\documentstyle[12pt,aasms4]{article}

\begin{document}

\title{The intermittent behavior and hierarchical clustering of the cosmic
mass field}

\author{Long-Long
Feng\altaffilmark{1}
\ Jes\'us Pando,\altaffilmark{2}
\ and
Li-Zhi Fang\altaffilmark{3}
}

\altaffiltext{1}{Center for Astrophysics, University of Science
and Technology of China, Hefei, Anhui 230026, National
Astronomical Observatories, Chinese Academy of Science, Chao-Yang
District, Beijing, 100012, P.R. China}
\altaffiltext{2}{Department of Chemistry and Physics, Chicago
State University, Chicago, IL 60628} \altaffiltext{3}{Department
of Physics, University of Arizona, Tucson, AZ 85721}

\begin{abstract}

The hierarchical clustering model of the cosmic mass field is
examined in the context of intermittency. We show that the mass
field satisfying the correlation hierarchy $\xi_n\simeq
Q_n(\xi_2)^{n-1}$ is intermittent if $\kappa < d$, where $d$ is
the dimension of the field, and $\kappa$ is the power-law index of
the non-linear power spectrum in the discrete wavelet transform
(DWT) representation.  We also find that a field with singular
clustering can be described by hierarchical clustering models with
scale-dependent coefficients $Q_n$ and that this scale-dependence
is completely determined by the intermittent exponent and
$\kappa$. Moreover, the singular exponents of a field can be
calculated by the asymptotic behavior of $Q_n$ when $n$ is large.
Applying this result to the transmitted flux of HS1700 Ly$\alpha$
forests, we find that the underlying mass field of the Ly$\alpha$
forests is significantly intermittent. On physical scales less
than about 2.0 h$^{-1}$ Mpc, the observed intermittent behavior
is qualitatively different from the prediction of the
hierarchical clustering with constant $Q_n$. The observations,
however, do show the existence of an asymptotic value for the
singular exponents. Therefore, the mass field can be described by
the hierarchical clustering model with scale-dependent $Q_n$. The
singular exponent indicates that the cosmic mass field at redshift
$\sim 2$ is weakly singular at least on physical scales as small
as 10 h$^{-1}$ kpc.

\end{abstract}


\section{Introduction}

The cosmic mass density field seems to posses two basic features
which at first glance appear contradictory. The first is the
statistical homogeneity and isotropy of the field. The second is
that the mass field consists of isolated density peaks like
galaxies and clusters of galaxies, in which the mass densities
are much higher than average. Moreover, coherent structures, like
filaments, sheets and even cellular or network structures with
characteristic lengths have been detected in galaxy distributions.

These two features can be reconciled by assuming the formation of
a highly intermittent field from an initially gaussian random
field. In this scenario, the growth of structure in the mass
field is initially linear and randomness is the mechanism by
which structure is formed.  At later stages, non-linearities
enter the dynamics and prevent infinite growth. As a result, the
structures that thereby arise in the random mass field have an
interesting character: they are strong enhancements (peaks) of
the density field scattered in a space with a low density
background. This feature generally is called intermittency and
was originally introduced for describing the temperature and
velocity distributions in turbulence (Batchelor \& Townsend 1949;
for review, see Frisch, 1995.) An early attempt of studying the
formation of the large scale coherent structures from a random
field is the theory of pancakes (Zeldovich 1970). This work is
the predecessor of the intermittency approach to the formation of
coherent structures (Shandarin \& Zeldovich 1984; for a review,
see Shandarin \& Zeldovich 1989).

The intermittency picture is supported by modelling the cosmic
matter distribution in non-linear regime with lognormal(LN)
random field, which is a simplest model of intermittency behavior
in random fields (e.g. Zeldovich, Ruzmaikin, \& Sokoloff 1990).
The LN model has been found to reporoduce the nonlinear matter
distribution evolved from Gaussian initial conditions and gives
rise to spatial patterns characterized statistically by a kind of
intermittency ('heterotopicity') (Cole and Jones, 1991; Jones,
Coles \& Martinez, 1992). The LN model of the baryonic matter
distribution (Bi \& Davidsen, 1997) has also been successfully
applied to explain the non-Gaussian features of the underlying
mass field traced by Ly$\alpha$ forests (Feng \& Fang 2000).

The intermittency picture is also supported by the singular
density profile of massive halos (like $\rho (r)\propto
r^{-\gamma}$ with exponent $\gamma \simeq 1$) given by N-body
simulations (e.g. Navarro, Frenk \& White, 1997, Jing \& Suto
2000.) This profile indicates that the events of large
differences between the local densities separated by small
distances $r$ are more frequent than with a gaussian field. As
such the probability distribution function(PDF) of the density
difference, $\Delta_r({\bf x})\equiv \rho({\bf x+r})-\rho({\bf
x})$, on small scales $r$ is long tailed.

The PDF of $\Delta_r({\bf x})$ can only be effectively detected
by a scale-space decomposition because $\Delta_r({\bf x})$
consists of both scale $r$ and position ${\bf x}$. A scale-space
decomposition analysis of the mass density field traced by the
Ly$\alpha$ forests has indeed obtained more direct evidence of
the intermittency (Jamkhedkar, Zhan, \& Fang 2000.) They found
that the PDF of $\Delta_r({\bf x})$ is gaussian on scale $r$
larger than about 2 h$^{-1}$ Mpc, but becomes longed tailed as $r$
decreases. Hence, the random mass field traced by the Ly$\alpha$
transmitted flux shows an excess of large fluctuation on small
scales in comparison to a Gaussian distribution. This is typical
of an intermittent field (Shraiman \& Siggia 2000.)

It seems that intermittency is a common feature of the
non-linearly evolved cosmic mass field. At the same time,
hierarchical clustering is believed to be a common feature of the
non-linearly evolved cosmic mass field, and has been widely
applied to construct semi-analytic models of gravitational
clustering in the universe (e.g. White 1979). Although
hierarchical clustered fields show indeed some kind of
intermittency, it is unclear whether the correlation hierarchy is
consistent with the observed intermittency of the cosmic mass
field. For instance, in the LN model, the high order correlations
do not satisfy a hierarchical relation, but obey a well-known
Kirkwood scaling relation (Kirkwood, 1935; Peebles, 1980).
Therefore, a LN random field is intermittent, but does not
satisfy the hierarchical clustering.

The purpose of this paper is to investigate the relationship
between intermittency and hierarchical clustering. We will not
limit ourselves to special models of intermittency, but will try
to calculate the intermittent feature of a hierarchical clustered
field in general. We first introduce the intermittent exponent
which gives a detailed classification of the non-linear
clustering. Specifically it gives a complete and uniform
description of the nonlinearity of the cosmic mass field from
initially gaussian perturbations, to weak, and eventually strong
non-linear clustering. We then study the intermittent exponent
predicted by the hierarchical clustering. As expected, the
intermittent behavior given by the hierarchical clustering model
is found to be in good agreement with observed results on large
scales. However, for the highly non-linearly evolved field, we
find that the simplest correlation hierarchy, i.e., one with
constant coefficients $Q_n$, is no longer adequate to describe the
observed features in the density field. In this case, the model
of hierarchical clustering with scale dependent $Q_n$ may still
work. Moreover, we find the relationship between the intermittent
and singular behaviors of the cosmic mass field and the scale
dependence of the $Q_n$.

The paper will be organized as follows. \S 2 introduces the basic
statistical measures of the intermittency of random mass fields --
the structure function and intermittent exponent. \S 3 presents
the predicted intermittency of a hierarchically clustered field.
The singular behavior of the hierarchical clustering with
scale-dependent $Q_n$ will be discussed in \S 4. The comparison
of the intermittency and singular behaviors of a hierarchically
clustered field with samples of QSO's Ly-$\alpha$ forests is
given in \S 5. Finally, the conclusions and discussions are in \S
6.

\section{Intermittency}

The chief characteristic of an intermittent field is that the
interesting stuff happens in a local area. As a consequence, the
description of intermittent fields by traditional methods such as
the Fourier power spectrum are ineffective because these methods
are essentially non-local. For instance, the amplitudes of the
Fourier coefficients lose the spatial information.  Two fields
which have a similar gaussian PDF of the amplitudes of the
Fourier coefficient may have very different intermittent
features. Though the phases of the Fourier coefficients do
contain the spatial information, they are not computationally
convenient.

The density profiles described in \S 1 look to be a good way to
the measure mass density peaks. The profiles are, however, only a
measure for the individual massive halos and not a statistical
measure of the entire random field consisting of these peaks. For
example, imagine a homogeneous gaussian field in a cubic box with
128$^3$ grids, which corresponds to 128$^3$ realizations of a
gaussian random variable. Rare peaks as high as 4$\sigma$ will
occur at some points. Obviously, the individual density profiles
are useless in measuring the excess of the large density
fluctuations with respect to a gaussian distribution. Moreover,
the exponent $\gamma$ introduced with the individual density
profiles $\rho (r)\propto r^{-\gamma}$ cannot be used to describe
a statistically homogeneous and isotropic field, which requires
that proper measures be statistically invariant quantities under
translational and rotational transformations.

The proper measure of an intermittent field should satisfy the
following conditions simultaneously: 1. its second order
statistics should describe the power spectrum of the field, and
2. the correlations of phases should describe coherent
structures. We show in this section that the structure functions
in the discrete wavelet transform (DWT) representation meet these
conditions.

\subsection{The local density difference}

The basic quantity for describing the intermittency of a random
density field $\rho({\bf x})$ is the {\it difference} between the
densities at the position ${\bf x}$ with separation ${\bf r}$,
i.e.,
\begin{equation}
\Delta_{ r}({\bf x})\equiv \rho({\bf x+r})-\rho({\bf x}) =
\frac{1}{\bar{\rho}} [\delta({\bf x+r})-\delta({\bf x})],
\end{equation}
where $\delta({\bf x})=[\rho({\bf x})-\bar{\rho}]/\bar{\rho}$ is
the density contrast. For simplicity, we assume that the density
is normalized, $\bar{\rho}=1$. The density difference is
essential to describe the most important intermittent
characteristics of a mass field. These features include:

1. {\em The power spectrum and the local power spectrum.}

   The ensemble average of the second moment of
$\Delta_r({\bf x})$ is
\begin{equation}
S^2(r) =\langle |\Delta_{r}({\bf x})|^2 \rangle.
\end{equation}
If the field is homogeneous, $S^2(r)$ is independent of ${\bf x}$
and depends only on $r$. $S^2(r)$ is the mean power of the density
fluctuations at wavenumber $k\simeq 2\pi/r$, and therefore,
$S^2(r)$ is the power spectrum of the field (see \S 2.3). For a
given realization of the random field, $|\Delta_{r}({\bf x})|^2$
is the {\em local} power spectrum at ${\bf x}$. For an
intermittent field the spatial distribution of $|\Delta_{r}({\bf
x})|^2$ is highly irregular or spiky. The power of the density
fluctuations on small $r$ is highly localized in space with very
low powers between these spikes. This is probably the easiest way
of identifying intermittency (Jamkhedkar, Zhan, \& Fang 2000.)

2. {\em The long-tailed PDF of} $\Delta_{r}({\bf x})$.

One of the defining characteristics of an intermittent field is
the long-tailed PDF of $\Delta_{r}({\bf x})$ on small scales $r$.
This trait can effectively be measured by the higher order moments
of $\Delta_r({\bf x})$,
\begin{equation}
S^{2n}(r) = \langle|\Delta_{r}({\bf x})|^{2n} \rangle
\end{equation}
where $n$ is a positive integer. The $S^{2n}(r)$ are called
structure functions. When the ``fair sample hypothesis" is
applicable (Peebles 1980), $S^{2n}(r)$ can be calculated by the
spatial average,
\begin{equation}
S^{2n}(r) =\frac{1}{V}\int |\Delta_{r}({\bf x})|^{2n}d{\bf x},
\end{equation}
where $V$ is the spatial normalization. When $n$ is large,
$S^{2n}(r)$ is dominated by the long-tail events.  Eqs. (2) and
(4) show that the structure function $S^{2n}_j$ unifies the
analysis of intermittency ($n > 1$) with the power spectrum
($n=1$).

3. {\em Density peaks and scale-scale correlations.}

The density profile of  mass halos $\rho (|{\bf x-x_0}|)\propto
|{\bf x-x_0}|^{-\gamma}$ at ${\bf x_0}$ means that events with
large $|\Delta_{r}({\bf x})|$ on different scales $r$ occur at
the same place ${\bf x=x_0}$. In other words, the density peaks
lead to the correlation between $|\Delta_{r}({\bf x})|$ and
$|\Delta_{r'}({\bf x})|$ with $r \neq r'$. As such, one can
distinguish the field containing only peaks caused by gaussian
fluctuations from an intermittent field by the scale-scale
correlations defined as (Pando et al. 1998)
\begin{equation}
C_{ r, r'}^{n,n'}=\frac{\langle
  \Delta_{r}({\bf x})^{n}\Delta_{r'}({\bf x})^{n'}\rangle}
{\langle \Delta_{\bf r}({\bf x})^{n}\rangle
  \langle \Delta_{\bf r'}({\bf x})^{n'}\rangle},
\end{equation}
where $n$ and $n'$ are even integers. $C_{\bf r, r'}^{n,n'}= 1$
for gaussian fields, while $C_{\bf r, r'}^{n,n'}> 1$ for
intermittent fields. When $n$ and $n'$ are large, $C_{r,
r'}^{n,n'}$ is dominated by the high density peaks. The excess of
high peaks (as compared to gaussian peaks) can be measured by the
scale-scale correlations.

4. {\em Singular behavior and H\"older exponent $\alpha$}

In the context of random fields, the singular behavior is
characterized by the H\"older exponent $\alpha$ defined as (Adler
1981)
\begin{equation}
\Delta_{r}({\bf x})= |\delta({\bf x +r})-\delta({\bf x})| \leq
(r/L)^{\alpha} \hspace{5mm} {\rm as} \ \ r \rightarrow 0,
\end{equation}
where the constant $L$ can be taken as the sample size. The
exponent $\alpha$ measures the smoothness of the field: for larger
$\alpha$, the field is smoother on smaller scales and vice versa.
If $\alpha$ is negative, the field becomes singular. In this case
the H\"older exponent is dominated by the events with extremely
large density difference $|\Delta_{r}({\bf x})|$ on extremely
small scales ($r \rightarrow 0$), and therefore, the exponent
$\alpha$ should be dependent on the index $\gamma$ of the singular
density profile if the object with the singular density profile
is typical of the mass field. For a homogeneous and isotropic
field, the PDF of the local density (contrast) difference
$[\delta({\bf x +r})-\delta({\bf x})]$ for a given scale $r$ has
to be independent of position ${\bf x}$. Therefore, the exponent
$\alpha$ defined by eq.(6) is statistically invariant under
translation and rotational.

Since the resolution of real and simulation samples is always
finite, one cannot practically measure the individual singularity
of $|{\bf x_0+r}|^{-\gamma}$ at position ${\bf x_0}$ with $r
\simeq 0$. However, $\alpha$ defined by eq.(6) can be calculated
as the asymptotic behavior of the statistics of local density
differences when $1/r \rightarrow \infty$. Hence, one can measure
the singular exponent $\alpha$ as a statistical property of the
random field, rather than using the density profiles of the
individual massive halos.

We should make an important distinction between statistics based
on $\Delta_{r} ({\bf x})$ and statistics that use the density
$\rho({\bf x})$ or density contrast $\delta({\bf x})$. The former
is the {\em difference} in either the density $\rho({\bf x})$ or
the density contrast $\delta({\bf x})$ at different positions in
space.  It is possible that a large difference $\Delta_{r} ({\bf
x})$ on small scales occurs in regions of low density, and also
possible that a small difference occurs in regions of high
density. To study intermittency, data which samples the entire
cosmic mass field is necessary.

Finally, an advantage of using the local density difference is in
reducing the contamination of the random velocity field. It is
well known that the random velocity field will smooth the mass
field in redshift space and repress the power of the density
perturbations on scales below that characterized by the velocity
dispersion. Because the difference $\Delta_{r}({\bf x})$ is
localized in space and scale, the velocity fluctuations on scales
larger than $r$ will lead to offset of the event $\Delta_{r}({\bf
x})$ in redshift space, but keep the magnitude of
$\Delta_{r}({\bf x})$ unchanged. As a consequence, the PDF of
$\Delta_{r}({\bf x})$ will not be affected by velocity
fluctuations on scales larger than $r$, and the spiky features of
the local power spectrum of the Ly$\alpha$ forests are still
significant on scales of a few tens h$^{-1}$ kpc (Jamkhedkar,
Zhan, \& Fang 2000.)

\subsection{The Intermittent exponent}

In the previous section we showed that the intermittency is
described by the scale-dependence of the higher order moments of
$\Delta_r({\bf x})$. In this section, we introduce the
intermittent exponent, which is a more effective tool of
describing this scale-dependence.

For a homogeneous and isotropic gaussian field, the structure
function eq.(3) can be calculated by
\begin{eqnarray}
S^{2n}(r)  & = & \int_{-\infty}^{\infty}
P_{g}(\Delta_r)[\Delta_r]^{2n} d\Delta_r \\ \nonumber
 & = & (2n-1)!!
   \left [\sigma_{\rho}^2(r)\right ]^n,
\end{eqnarray}
where $P_{g}(\Delta_r)$ is the gaussian PDF of $\Delta_r({\bf
x})$, and $\sigma^2(r)=\langle |\Delta_r|^{2}\rangle$ is the
variance of $P_{g}(\Delta_r)$.

The deviation of the field from a gaussian distribution on scale
$r$ can be measured by the intermittent exponent defined
as\footnote{In turbulence, $\bar{S}^{2n}(r)$ is used to define
the so-called anomalous scaling describing the intermittent
behavior (Shraiman \& Siggia 2000.)}
\begin{equation}
\bar{S}^{2n}(r) \equiv \frac{S^{2n}(r)}{[S^{2}(r)]^n} \propto
   \left(\frac{r}{L} \right )^{\zeta}.
\end{equation}
$\zeta$ is equal to zero if the mass density field is gaussian on
all scales. From eq.(8), we have
\begin{equation}
\frac{\bar{S}^{2n}(r)}{\bar{S}^{2n}(r_0)}=
   \left(\frac{r}{r_0} \right )^{\zeta},
\end{equation}
where again, $\zeta$ is equal to zero if the mass density field is
gaussian on all scales.

A field is self-similar if the difference $\Delta_{r}({\bf x})$ as
a random variable satisfies
\begin{equation}
\Delta_{r}({\bf x}) = \lambda^{h}
 \Delta_{\lambda r}({\bf x}),
\end{equation}
where $h$ is constant. In this case, $\bar{S}^{2n}(r)$ is
independent of $r$. Once again, the exponent $\zeta$ defined by
eq.(9) is zero. A field is said to be intermittent if the
exponent $\zeta$ is non-zero. Hence, {\it an intermittent field
is neither gaussian nor selfsimilar} (Frisch, 1995.)

If the field contains more ``abnormal'' events, i.e., large
$\Delta_{r}({\bf x})$ on small scales, $\bar{S}^{2n}(r)$ will be
dominated by these events, especially when $n$ is large. In this
case, we have $S^{2n}(r) >[{S^{2}(r)}]^n$ on small $r$, and
therefore, the exponent $\zeta$ is less than zero. Similar to the
exponent $\alpha$ [eq.(6)], the intermittent exponent $\zeta$
measures the smoothness of the field: for larger $\zeta$, the
field is smoother on smaller scales, and vice versa. If $\zeta$
is negative on small scales, the field is probably singular.

The $n$- and $r$-dependence of $\zeta$ provide a detailed
description of the non-linear features of the mass field. $\zeta$
gives an unified criterion for classifying the non-linear
features of a random field, from gaussian, to self-similar, to
mono- and multi-fractal, and to singular.

\subsection{The Intermittent exponent in the DWT representation}

The density contrast difference $\Delta_r({\bf x})$ [eq.(1)]
contains two variables: the position ${\bf x}$ and the scale $r$,
and therefore, the information of $\Delta_r({\bf x})$ can best be
extracted by a proper space-scale decomposition.

We will use the discrete wavelet transform (DWT) decomposition
[For details on the DWT refer to Mallat (1989a,b); Meyer (1992);
Daubechies, (1992), and for physical applications, refer to Fang
\& Thews (1998)]. To simplify the notation, we consider only the
DWT decomposition for 1-D density field $\rho(x)$ or $\delta(x)$
on a spatial range $L$. It is straightforward to generalize the
result to 2- and 3-D fields.

We first divide the spatial range $L$ into $2^j$ segments labelled
by $l=0,1 ..2^j-1$. Each of the segments has length $L/2^j$. The
density contrast difference $\Delta_{r}(x)=\delta(x+r)-\delta(x)$
defined in eq.(1) can be replaced by
\begin{equation}
\tilde{\epsilon}_{j,l} =\sqrt{\frac{2^j}{L}}
  \left [\int_{lL/2^j}^{(l+1/2)L/2^j}\delta(x)dx
   -\int_{(l+1/2)L/2^j}^{(l+1)L/2^j}\delta(x)dx \right ].
\end{equation}
$\tilde{\epsilon}_{j,l}$ measures the difference between the mean
density contrasts in the localized segments $lL/2^j \leq x <
(l+1/2)L/2^j$ and $(l+1/2)L/2^j \leq x < (l+1)L/2^j$. Therefore,
it can be identified as $\Delta_r(x)$ with $x = lL/2^j$ and $r
=L/2^j$.

$\tilde{\epsilon}_{j,l}$ is the Haar wavelet function coefficient
(WFC), given by the projection of $\delta(x)$ onto the Haar
wavelet basis $\psi_{j,l}^{H}(x)$ as
\begin{equation}
\tilde{\epsilon}_{j,l}= \int \delta(x)\psi_{j,l}^{H}(x)dx,
\end{equation}
where
\begin{equation}
\psi^{H}_{j,l}(x) =\sqrt{\frac{2^{j}}{L}} \left\{
\begin{array}{ll}
  1 & \mbox{from $Ll2^{-j}$ to $Ll2^{-j}+L2^{-j-1}$} \\
 -1 & \mbox{from $Ll2^{-j}+L2^{-j-1}$ to $L(l + 1)2^{-j}$} \\
  0 &
\mbox{otherwise.}
\end{array} \right.
\end{equation}
The factor $\sqrt{2^{j}/L}$ insures the orthonormality of the
wavelet with respect to both indices $j$ and $l$.

For other wavelets $\psi_{j,l}(x)$, the WFC,
$\tilde{\epsilon}_{j,l}$, also measures the difference of the
density contrast on separation  $L/2^j$ at a position $lL/2^j$.
We will use the Daubechies 4 wavelet (Daubechies, 1992) in our
numerical calculations below because it is better behaved in
scale space than the Haar wavelet (Fang \& Thews 1998). The set
of wavelets $\psi_{j,l}(x)$ ($j=0,1,..$ and $l=0,1,..2^j -1$) is
complete and orthogonal. In the DWT representation, the field
$\delta(x)$ can be expressed as
\begin{equation}
\delta(x) =
     \sum_{j'=0}^{J} \sum_{l=0}^{2^{j'}-1} \tilde{\epsilon}_{j',l}
 \psi_{j',l}(x).
\end{equation}
$J$ is determined by the spatial resolution $\Delta x$ of the
sample, i.e., $J \simeq \log_2(L/\Delta x)$. The WFC
$\tilde{\epsilon}_{j',l}$ is given by eq.(12), replacing the Haar
wavelet $\psi_{j,l}^{H}(x)$ by $\psi_{j,l}(x)$. Since wavelets
are admissible, the WFC can be calculated with either the density
or the density contrast
\begin{equation}
\tilde{\epsilon}_{j,l}=\int \rho(x)\psi_{j,l}(x)dx =
\frac{1}{\bar{\rho}} \int \delta(x)\psi_{j,l}(x)dx.
\end{equation}

In the DWT representation, eq.(2) is
\begin{equation}
S^2_j =\frac{1}{2^j}\sum_{l=0}^{2^j-1}|\tilde{\epsilon}_{j,l}|^2.
\end{equation}
This is the power spectrum in the DWT representation and equal to
a band-averaged Fourier power spectrum given by (Fang \& Feng
2000)
\begin{equation}
S^2_j = \frac{1}{2^j} \sum_{n = - \infty}^{\infty}
 |\hat{\psi}(n/2^j)|^2 P(k),
\end{equation}
where $P(k)$ is the Fourier power spectrum with the wavenumber
$k=2\pi n/L$.  $\hat{\psi}(n)$ is the Fourier transform of the
basic wavelet $\psi(\eta)$.

The structure function (3) in the DWT representation is
\begin{equation}
S^{2n}_j = \langle|\tilde{\epsilon}_{j,l}|^{2n} \rangle
=\frac{1}{2^j}\sum_{l=0}^{2^j-1}|\tilde{\epsilon}_{j,l}|^{2n}
\end{equation}
and eq. (8) is
\begin{equation}
\bar{S}^{2n}_j=\frac{S^{2n}_j}{[S^2_j]^n} \propto 2^{-j\zeta}.
\end{equation}
It has been shown that $\zeta$ given by eq.(19) is the same as (8)
(Jaffred 1994.) $\bar{S}^{2n}_j$ or $\zeta$ is the basic measure
of the intermittency of the cosmic mass fields.

\section{Intermittency within hierarchical clustering}

Hierarchical clustering provides a model describing the non-linear
clustering of the cosmic mass field. It assumes that the
correlation functions of the mass density can be described by the
linked-pair approximation, i.e., the $n$-th irreducible
correlation function $\xi_n$ is given by the two-point
correlation function $\xi_2$ as $\xi_n = Q_n \xi_2^{n-1}$, where
$Q_n$ is the hierarchical coefficient (White 1979). As the high
order correlation functions provide a direct way to investigate
the nonlinearity in a random field and test the scaling
hierarchy, it follows that one can study the hierarchical
clustering predictions pertaining to intermittency and the
singular behavior of the mass field. In this section we develop
the higher order correlation functions in the DWT representation.

\subsection{Linked-pair approximation with the DWT modes}

For the $n$-th order correlation function, the correlation
hierarchy, or linked pair approximation, is
\begin{equation}
\xi_n({\bf x_1,..x_n}) =
\sum_{\varpi}Q_n^{\varpi}\sum_{(ab)}\prod^{n-1}
  \xi_2(x_{ab}),
\end{equation}
where coefficients $Q_n^{\varpi}$ are constant, the sum $\varpi$
is over the types of the tree graphs with $n$ vertices,
$x_{ab}=|{\bf x_a-x_b}|$, and the sum $(ab)$ is over relabelings
within $\varpi$ (Fry 1984). With this relation, the higher order
behavior of the clustering is completely determined by the
two-point correlation $\xi_2(x_{ab})$ and the constants
$Q_n^{\varpi}$. Hence, the intermittent behavior of a
hierarchical clustered field can be determined by the
coefficients $Q^{\varpi}_n$ and the non-linear power spectrum.

For instance, in the case of $n=4$, eq.(20) yields
\begin{eqnarray}
\lefteqn{\langle \delta({\bf x_1}) \delta({\bf x_2}) \delta({\bf
x_3}) \delta({\bf x_4})\rangle  =  } \\ \nonumber
 & &  Q^a_4[\langle \delta({\bf x_1}) \delta({\bf x_2})\rangle
\langle \delta({\bf x_2}) \delta({\bf x_3})\rangle \langle
\delta({\bf x_3}) \delta({\bf x_4})\rangle +
 {\rm cyc. \ 11 \ terms}]\\ \nonumber
  & & +Q^b_4[\langle \delta({\bf x_1}) \delta({\bf x_2})\rangle
\langle \delta({\bf x_1}) \delta({\bf x_3})\rangle \langle
\delta({\bf x_1}) \delta({\bf x_4})\rangle + {\rm cyc. \ 3 \
terms}].
\end{eqnarray}
where $Q^a_4$ is for snake diagrams, and $Q^b_4$ is for stars.

We express eq.(21) in the DWT basis, $\psi_{\bf j,l}({\bf
x_1})\psi_{\bf j,l}({\bf x_2})
  \psi_{\bf j,l}({\bf x_3})\psi_{\bf j,l}({\bf x_4})$, where
$\psi_{\bf j,l}({\bf x})$ is the 3-D wavelet (${\bf
x}=x^1,x^2,x^3)$. $\psi_{\bf j,l}({\bf x})$ is given by a direct
product of the 1-D wavelets, i.e., $\psi_{\bf j,l}({\bf
x})=\psi_{j_1,l_1}(x^1) \psi_{j_2,l_2}(x^2) \psi_{j_3,l_3}(x^3).$

Equation (21) yields
\begin{eqnarray}
\lefteqn{\langle \tilde{\epsilon}_{\bf j,l}^4\rangle
   = Q^a_4[\sum_{\bf j',l'}\sum_{\bf j'',l''}\sum_{\bf j''',l'''}
     \sum_{\bf j'''',l''''}
  \langle \tilde{\epsilon}_{\bf j,l}\tilde{\epsilon}_{\bf j',l'}\rangle
\langle \tilde{\epsilon}_{\bf j'',l''}\tilde{\epsilon}_{\bf
j''',l'''} \rangle \langle \tilde{\epsilon}_{\bf
j'''',l''''}\tilde{\epsilon}_{\bf j,l}\rangle }
  \\ \nonumber
 & & \int\psi_{\bf j,l}({\bf x_2})\psi_{\bf j',l'}({\bf x_2})
     \psi_{\bf j'',l''}({\bf x_2})d{\bf x_2}
\int\psi_{\bf j,l}({\bf x_3})\psi_{\bf j''',l'''}
 ({\bf x_3})\psi_{\bf j'''',l''''}({\bf x_3})d{\bf x_3}
  \\ \nonumber
& & + {\rm cyc. \ 11 \ terms}]  \\ \nonumber
  & & + Q^b_4[\sum_{\bf j',l'}\sum_{\bf j'',l''}\sum_{\bf j''',l'''}
\langle\tilde{\epsilon}_{\bf j',l'}\tilde{\epsilon}_{\bf
j,l}\rangle \langle\tilde{\epsilon}_{\bf
j'',l''}\tilde{\epsilon}_{\bf j,l}\rangle
\langle\tilde{\epsilon}_{\bf j''',l'''}\tilde{\epsilon}_{\bf
j,l}\rangle
    \\ \nonumber
 & & \int\psi_{\bf j,l}({\bf x_1})\psi_{\bf j',l'}({\bf x_1})
   \psi_{\bf j'',l''}({\bf x_1})\psi_{\bf j''',l'''}({\bf x_1})d{\bf x_1}
   + {\rm cyc. \ 3 \ terms}]
\end{eqnarray}
where the {\bf j} and {\bf l} terms are the 3-D scales and
positions respectively.

The DWT is very efficient in compressing data, i.e., the off
$j$-diagonal elements of the covariance $\langle
\tilde{\epsilon}_{j,l}\tilde{\epsilon}_{j',l'}\rangle$ generally
are much less than the $j$-diagonal elements. This property is
also consistent with the quasi-locality of mass clustering
(Pando, Feng \& Fang 2001.) Thus, the r.h.s. of eq.(22) is
dominated by the terms $\langle \tilde{\epsilon}_{\bf j,l}^2
\rangle^{3}$ and we have
\begin{equation}
\langle \tilde{\epsilon}_{\bf j,l}^4 \rangle \simeq [Q^a_4A_{\bf
j}+ Q^b_4 B_{\bf j}]
  \langle \tilde{\epsilon}_{\bf j,l}^2 \rangle^{3},
\end{equation}
where the factors $A_j$ and $B_j$ are given by
\begin{equation}
A_{\bf j}=\left[\int \psi_{\bf j,l}^3({\bf x})d{\bf x}\right]^2
\end{equation}
\begin{equation}
B_{\bf j}=\int \psi_{\bf j,l}^4({\bf x})d{\bf x}.
\end{equation}

Because wavelets are constructed by dilating and translating a
basic wavelet $\psi(\eta)$, $A_{{\bf j}}$ and $B_{{\bf j}}$ can
be more simply written as
\begin{equation}
A_{\bf j}=2^{j^1+j^2+j^3}\frac{1}{L^3}
             \left[\int \psi^3({\bf \eta})d{\bf \eta}\right]^2
\end{equation}
and
\begin{equation}
B_{\bf j}=2^{j^1+j^2+j^3}\frac{1}{L^3}
 \int \psi^4({\bf \eta})d{\bf \eta}
\end{equation}
Thus, eq.(23) becomes finally
\begin{equation}
\langle \tilde{\epsilon}_{\bf j,l}^4 \rangle \simeq Q_4
2^{j^1+j^2+j^3}
  \langle \tilde{\epsilon}_{\bf j,l}^2 \rangle^{3},
\end{equation}
where the coefficient $Q_4$ is independent of ${\bf j}$ and given
by
\begin{equation}
Q_4=\frac{1}{L^3} \left[ Q^a_4\left[\int \psi^3({\bf \eta})d{\bf
\eta}\right]^2 + Q^b_4\int \psi^4({\bf \eta})d{\bf \eta}\right ].
\end{equation}

Similarly, for $2n$-th order, we have
\begin{equation}
\langle \tilde{\epsilon}_{\bf j,l}^{2n}\rangle \simeq Q_{2n}
2^{(n-1)(j^1+j^2+j^3)}
  \langle \tilde{\epsilon}_{\bf j,l}^2 \rangle^{2n-1},
\end{equation}
where again $Q_{2n}$ is independent of ${\bf j}$. Eq.(30) is the
counterpart of correlation hierarchy $\xi_n=Q_n\xi^{n-1}_2$ in
the DWT representation.

For a 1-D field, such as Ly$\alpha$ forests, we use a projection
of a 3-D distribution $\delta({\bf x})$ onto 1-D as
\begin{equation}
\tilde{\epsilon}_{j,l}= \int_{-\infty}^{\infty} \delta({\bf
x})\psi_{j,l}(x^1)
  \phi_{J',m}(x^2)\phi_{J',n}(x^3)dx^1dx^2dx^3,
\end{equation}
where $x^1$ is the redshift , $x^2$ and $x^3$ are the positions
in the sky, and $\phi_{j,l}(x)$ is the scaling function of the
DWT analysis. The scaling function  $\phi_{j,l}(x)$ plays the
role of a window function on scale $j$ at position $l$ (Fang and
Thews 1998). With eq.(31), the 1-D field linked-pair
approximation (20) yields
\begin{equation}
\langle \tilde{\epsilon}_{j,l}^{2n}\rangle \simeq Q_{2n}
2^{j(n-1)}
  \langle \tilde{\epsilon}_{j,l}^2 \rangle^{2n-1},
\end{equation}
where $Q_{2n}$ is for a 1-D field and is generally not equal to
$Q_{2n}$ for a 3-D field because the difference in the projection
(31) for the 1-D case. We use the same notation $Q_{2n}$ for both
1-D and 3-D and from the context it will be clear whether we are
using the 1-D or 3-D quantity.

\subsection{The intermittency exponent within hierarchical clustering}

Applying eq.(30) for the diagonal case, i.e. $j^1=j^2=j^3=j$, we
have
\begin{equation}
\langle \tilde{\epsilon}_{\bf j,l}^{2n}\rangle \simeq Q_{2n}
2^{3j(n-1)}
  \langle \tilde{\epsilon}_{\bf j,l}^2 \rangle^{2n-1}.
\end{equation}
One can rewrite eqs.(32) and (33) as
\begin{equation}
\langle \tilde{\epsilon}_{\bf j,l}^{2n}\rangle \simeq Q_{2n}
2^{dj(n-1)}
  \langle \tilde{\epsilon}_{\bf j,l}^2 \rangle^{2n-1},
\end{equation}
where $d$ is the number of dimensions.

Substituting eq.(34) into eq.(19), we have
\begin{equation}
\bar{S}^{2n}_{\bf j}=Q_{2n} 2^{dj(n-1)}
  \langle \tilde{\epsilon}_{\bf j,l}^2 \rangle^{n-1}.
\end{equation}
Therefore, the intermittent exponent of a hierarchical clustered
field is
\begin{equation}
\zeta_h= - (n-1)\left [d + \frac{1}{j}\ln_2 P_j \right ] -
  \frac{1}{j}\ln_2 Q_{2n},
\end{equation}
the subscript $h$ is used to emphasize that $\zeta_h$ is the
predication in the hierarchical clustering model. For small
scales (large $j$), the last term of eq.(36) is negligible if the
$Q_{2n}$'s are constant. Hence, for a hierarchically clustered
field, the intermittency exponent $\zeta_h$ is completely
determined by the power spectrum $P_j=\langle
\tilde{\epsilon}_{j,l}^2 \rangle$ in the non-linear regime. If
the power spectrum is a power law, $P_j\propto 2^{-j\kappa}$, we
have
\begin{equation}
\zeta_h \simeq -(n-1)(d-\kappa).
\end{equation}
Hence, the field is intermittent if $\kappa < d $.

The intermittent exponent eq.(37) is independent of $j$. This is
the simplest type of intermittency -- monofractal with fractal
dimension $\kappa$. That is, the space filling of the density
perturbations will be less by a factor $(1/2)^{(d-\kappa)}$ from
scale $j$ to $j+1$,  and in effect, the hierarchical clustering
model with constant $Q_n$ is equivalent to assuming that the
random field $\delta({\bf x})$ is monofractal.  This conclusion
is consistent with the phenomenological model of the hierarchical
relations developed by Soneira \& Peebles (1977). The model is
essentially based on a dimension $\kappa$ fractal
distribution.\footnote{This model is also similar to the
so-called $\beta$-model of the intermittency of turbulence, for
which the exponent $\zeta$ has similar terms as eq.(42) (Frisch
1995.)}

\section{Singular behavior of hierarchical clustering}

\subsection{Singular exponent}

We now consider the singular exponent $\alpha$ of eq.(6). If the
mass field contains singular structures like $\rho\propto
r^{-\gamma}$ with $\gamma>0$ at position ${\bf x_0}$ (or ${\bf
l_0}$), the modulus of the diagonal WFCs ($j_1=j_2=j_3=j$) satisfy
\begin{equation}
|\tilde{\epsilon}_{j,l_0}|  \leq A 2^{j(\gamma - d/2)},
\end{equation}
where the factor $d/2$ is from the normalization $\sqrt{2^j/L}$ in
eq.(13). Eq.(38) shows that the singular exponent defined in
eq.(6) is $\alpha=\gamma - d/2$. Thus, one can measure $\gamma$
asymptotically as follows.

Because for large $n$, $S^{2n}_j$ is dominated by events with
large $\tilde{\epsilon}_{j,l}$, eq.(19) gives
\begin{equation}
\gamma \asymp \frac{d}{2} -\frac{1}{2n}\zeta + \frac{1}{2j}\ln_2
P_j,
 \ \ {\rm when} \ n \ {\rm is \ large}.
\end{equation}
Thus, a field is singular if the following condition holds
\begin{equation}
\frac{d}{2} -\frac{1}{2n}\zeta + \frac{1}{2j}\ln_2 P_j >0,
 \ \ {\rm when} \ n \ {\rm is \ large}.
\end{equation}

For a hierarchical clustered field, condition (40) is
\begin{equation}
  d-\kappa > 0  \ \ {\rm when} \ n \ {\rm is \ large}.
\end{equation}
Therefore, a hierarchical clustered field cannot be singular if
the power index $\kappa > d$. Under condition (41), eq.(37) leads
to $\zeta <0$. This result is consistent with the statement that
the field is singular when $\zeta <0$ on small scales (\S 2.2).

\subsection{Hierarchical clustering with scale dependent coefficients}

In order to have the hierarchical clustering scenario compatible
with observed data and N-body simulation samples, it is often
assumed that the $Q_n$ are scale dependent. In this section, we
will show that the scale dependence of the $Q_n$ is given by the
intermittent exponents. The validity of the hierarchical
clustering model with scale dependent $Q_n$ can be tested by
examining the relationship between the $Q_n$ scale dependence and
the singular exponents.

For the hierarchical clustering models with scale-dependent
$Q_{2n}$, eq.(34) still holds, as all the quantities in eq.(34)
are on  same scale $j$.  Thus, the scale-dependence of $Q_{2n}$
is given by
\begin{equation}
Q_{2n}= \frac{1}{2^{dj(n-1)}} \frac{ \bar{S}^{2n}_{\bf
j}}{P_j^{n-1}}.
\end{equation}
The scale dependence of $Q_{2n}$ can be measured by a power law
index defined by
\begin{equation}
Q_{2n} \propto 2^{j\beta_{2n}}.
\end{equation}
Using eqs.(19) and (42), eq.(43) yields
\begin{equation}
\beta_{2n} = -d(n-1) - \zeta - \frac{n-1}{j}\ln_2 P_j.
\end{equation}
We see that the scale-dependence of $Q_{2n}$ is completely
determined by the intermittent exponent $\zeta$ and the
non-linear power spectrum.

Using eq.(39), the singular exponent $\gamma$ can be expressed by
$\beta_{2n}$ as
\begin{equation}
\gamma \asymp d + \frac{1}{2n}\beta_{2n} +\frac{1}{j}\ln_2 P_j,
  \ \ \ {\rm when} \ n \ {\rm is \ large}.
\end{equation}
Eq. (45) shows that the singular index $\gamma$ is determined by
the scale-dependence of the coefficients $Q_{2n}$ and the
non-linear power spectrum $P_j$.

The important point of eq.(45) is that the l.h.s. is given by the
singular profiles such as $\rho \propto r^{-\gamma}$, independent
of $n$, while the r.h.s. consists of $n$-dependent quantities
like $\beta_{2n}$. Therefore, a test of the scale-dependent
$Q_{n}$ model is to check whether $\gamma$ given by eq.(45) is
$n$-independent when $n$ is large.

It is known that a scale-dependent intermittent exponent indicates
that the mass field is no longer monofractal, but is instead,
multifractal (Farge at al. 1996.) Galaxy distributions have been
shown to be multi-fractal in nature (Jones et.al, 1988) and to
explain this distribution, a phenomenological model of non-linear
fragmentation with multi-scaling parameters has been proposed
(Jones, Coles \& Martinez, 1992). Our present study has revealed
the intrinsic relationships between the scale-dependence of
$Q_{2n}$, multifractal nature, and singular behavior of the mass
field.

This completes our derivation of intermittency within hierarchical
clustering. With eq.(18) we pick up the DWT power spectrum,
eq.(37) gives the intermittency exponent predicted by the
hierarchical picture, eq.(42) provides a way to check for the
scale dependence of the correlation coefficients, and eq.(45)
gives a measure of singular behavior of the field. The
versatility of this approach is clear.

\section{Intermittent exponents of Ly$\alpha$ forests}

Since the intermittent and singular features of the cosmic mass
field are measured by the scale-dependence of the structure
functions, samples covering a large range of scales are best
suited for this analysis. As has been emphasized, the
intermittent and singular features are traits of the mass field
and not of the individual massive halos. As such, the transmitted
flux of QSO's Ly$\alpha$ forests given by high resolution
absorption spectrum are ideal data sets with which to work. The
Ly$\alpha$ transmitted flux is due to absorption by gases in cool
and low density regions. The pressure gradients are generally
less than the gravitational forces. The distribution of cool
baryonic diffuse matter is almost point-by-point proportional to
the underlying dark matter density (Bi, Ge \& Fang 1995). The
statistical features of the underlying mass field of the
Ly$\alpha$ forests can be detected by the transmitted flux.

\subsection{Samples}

We use the Ly$\alpha$ transmitted flux of QSO HS1700+64 ($z$ =
2.72) for our analysis. This sample has been employed to study
the baryonic matter density (Bi \& Davidsen 1997), the Fourier
and DWT power spectra and non-gaussian features (Feng \& Fang
2000). The recovered power spectrum has been found to be
consistent with the CDM cosmogony on scales larger than about 0.1
$h^{-1}$ Mpc. The data ranges from 3727.012\AA $ $ to 5523.554\AA,
for a total of 55882 pixels. On average, a pixel is about
0.028\AA, equivalent to a physical distance of $\sim 5$ h$^{-1}$
kpc at $z \sim 2$ for an Einstein-de Sitter universe. These
scales weakly depend on cosmological density parameter $\Omega$.
One can safely ignore the effect of the non-linear relation
between redshift and distance for our present analysis. In this
paper, we use the data from $\lambda =$ 3815.6\AA $ $ to
4434.3\AA, which corresponds to $z = 2.14 \sim 2.65$. The
comoving spatial size of our data is $189.84$ h$^{-1}$ Mpc in the
CDM model. This means for scale $j$, the spatial size is
$189.84/2^j$ h$^{-1}$ Mpc.

With Ly$\alpha$ data there is always the danger of contamination
from the presence of metal lines. We use three ways to estimate
the error this contamination causes. First, we block out the
significant metal line regions. Since the WFC
$\tilde{\epsilon}_{j,l}$ are localized, the metal line regions
are easily separated from the rest. Any effect from these regions
can be removed by not counting the DWT modes in the blocked
regions. The second way is to fill those regions with random data
which has the same mean power as the rest of the original data
and to smooth the data over the boundaries. Lastly we discard the
metal line chunks and smoothly connect the good chunks of data.
We find that the results not sensitive to the method of removing
metal lines.

Another source of contamination is noise. To estimate the effect
of noise we smooth the QSO's spectrum by filtering out all
extremely sharp spikes in the local power spectra on finest
scales. These spikes are caused by relatively strong fluctuations
between two neighboring pixels. Since such events are on the
smallest scales only, the analysis on larger scales does not
depend on whether we smooth the sample or not.

The DWT power spectrum of the Ly$\alpha$ transmitted flux of QSO
HS1700+64 is shown in Fig. 1. It is clear that $\kappa$ is larger
than 1 on all small scales $j \geq 9$ or physical scales $<$ 100
h$^{-1}$ kpc.

\subsection{Observed intermittent exponent}

We calculate the intermittent exponent with eq.(19),
\begin{equation}
\zeta = -\frac{1}{j}\ln_2\bar{S}^{2n}_j.
\end{equation}
The result is plotted in Fig. 2 and shows that on scales $j \geq
9$, the intermittent exponents are always significantly less than
zero. The underlying mass field of the QSO forests is neither
non-gaussian nor non-selfsimilar, but essentially intermittent on
physical scales less than 100 h$^{-1}$ kpc.

We next test the predicted relation (36) between the intermittent
exponent and power spectrum. The results of $\zeta_h$ are plotted
in Fig. 3. Because the power law index $\kappa$ of the DWT power
spectrum is generally larger than 1, $\zeta_h$ is always
positive. This is qualitatively different from $\zeta$ on small
scales. Therefore,  only on large scales $j \leq 5$ or physical
scales $>$ 2 h$^{-1}$ Mpc, does the correlation hierarchy with
constant $Q_n$ match observations.

\subsection{Multifractal scaling and singular exponent}

To check for hierarchical clustering with scale-dependent
coefficients, we calculate $Q_{2n}$ with eq.(42). The results are
shown in Fig. 4, which shows that only for lower $n$, can the
coefficient approximately be considered as constant. At all
higher $n$, $Q_{2n}$ significantly increases with decreasing
scale after $j>9$, which implies the breaking of a single scaling
hierarchy in the baryonic matter distribution.

Fig. 5 plots the singular exponent $\gamma$ calculated by eq.(45).
It indeed shows that for each $j$, $\gamma$ is asymptotically
independent of $n$, when $n$ is large. That is, models with
scale-dependent $Q_n$ are reasonable and the asymptotic values of
$\gamma$ are probably given by the exponent of the singular mass
density profile.

Fig. 5 also show that the singular exponents $\gamma$ become
larger with increasing scales. For instance, $\gamma \simeq 0.45$
at $j=9$, while $\gamma \simeq 0.18$ at $j=13$. This is
understandable if we consider the so-called  ``universal''
density profile of massive halos (Navarro, Frenk \& White 1997)
\begin{equation}
\rho(r)\sim \frac{1}{r(r+a)^2}
\end{equation}
where $a$ is the core radius. From this profile it is clear that
$\gamma$ on small scale $r < a$ will be less than for $r>a$.
Comparing with (47), the $\gamma$ given by the Ly$\alpha$
transmitted flux are small and implies that the singular behavior
of mass field at redshift $z \simeq 2$ is much weaker than the
present.

Although Fig. 5 shows that $\gamma >0$ on scale $j=13$ or
$\simeq$ 7 h$^{-1}$ kpc, the field is not necessarily singular.
For instance, if the profile (47) is modified as
\begin{equation}
\rho(r)\sim \frac{1}{(r+b)(r+a)^2}
\end{equation}
where physical scale $b \sim $ 5 h$^{-1}$ kpc, $\gamma$ will no
longer be positive on scales $j \geq 14 $.

\section{Conclusion and discussions}

The structure functions provide a complete and unified way of
studying intermittent fields. With this method, we have shown
that intermittency and hierarchical clustering in the non-linear
mass field are related. We find that if the field is singular,
the intermittent behavior on higher order $n$ can be completely
described by the singular exponents and the non-linear power
spectrum. In this case, the scale-dependencies of the
hierarchical clustering  coefficients $Q_{2n}$ at large $n$ are
also completely determined by these parameters.

The transmitted flux of HS1700 does show the existence of the
asymptotic value of the singular exponents. The cosmic mass field
at redshift $\sim 2$ is probably weakly singular at least on
physical scales as small as  10 h$^{-1}$ kpc. A singular field
can be neither self-similar nor mono-fractal. It is well known
that under many different conditions the field given by
gravitational clustering is self-similar. Therefore, the presence
of singular clustering implies a broken self-similarity and/or
mono-fractal.

 From a dynamical point of view, there is a basis for understanding
the relation between the intermittency and hierarchical
clustering. The basic dynamical processes of hierarchical
clustering are the merging of two halos or the accretion of small
halos into bigger ones. Regardless of specific processes, either
the infall of small halos into massive ones or the merging of
massive halos, these processes are stochastic. Thus, the
hierarchical clustering is described by equations containing the
addition and multiplication of stochastic variables. But these
stochastic processes are typical of the origin of intermittency
(e.g. Nakao 1998.). Even when the initial PDF of the random
variables is gaussian, the dynamics involving additive and
multiplicative stochastic variables will evolve to a long tailed
PDF. For instance, the growth of the mass $M$ of a halo is
sometimes phenomenologically described by the rate equation of
halo mass growth (e.g. Cole \& Lacey 1996, Salvador-Sol\'e,
Solanes \& Manrique 1998.) The rate equation is a
phenomenological model of the formation of intermittency via
merging.

The  so-called merging history tree theory has widely been
employed to describe the mean merging history of individual halos
and then to fit with the statistical properties of galaxies. Many
merging tree models have been proposed. In the context of
intermittency, different merging trees have different rules for
merging from smaller to larger scales, and therefore, lead to
different scale-scale correlations (\S 2.1). Thus, discrimination
among merging tree models is possible via a intermittent analysis.

\acknowledgments

We thank Dr. Wolung Lee for his support to this work. We also
thank Dr. D. Tytler for kindly providing the data of the Keck
spectrum HS1700+64. LLF acknowledges support from the National
Science Foundation of China (NSFC) and National Key Basic
Research Science Foundation.

\clearpage

\begin{figure}
\figurenum{1} \epsscale{0.6}
\plotfiddle{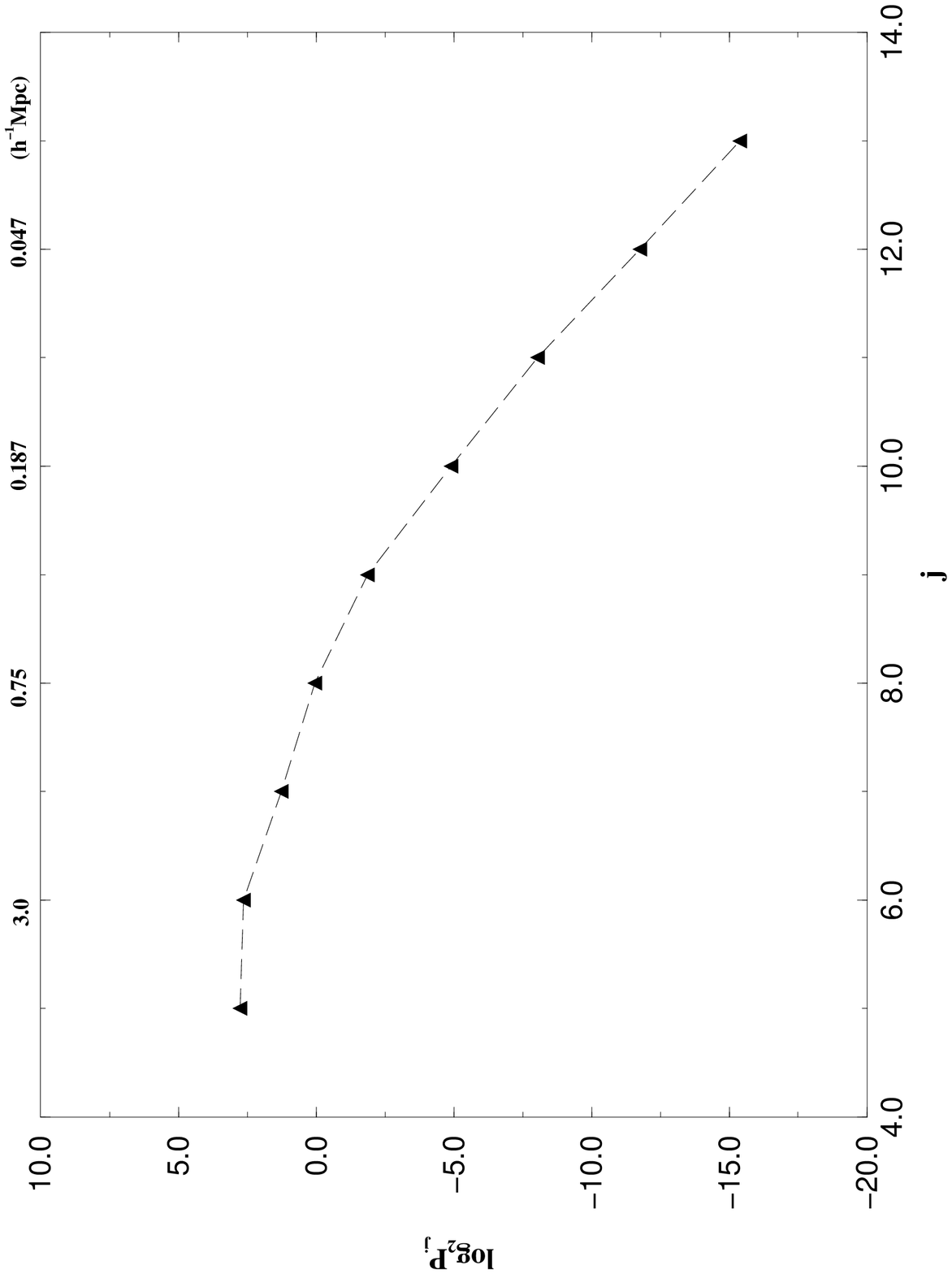}{20.cm}{270}{80}{80}{-310.}{440.}
\caption[]{The DWT power spectrum of the Ly$\alpha$ transmitted
flux of QSO HS1700+64. The top scale is the comoving scale, i.e.,
$189.84/2^j$ h$^{-1}$ Mpc.}
\end{figure}

\begin{figure}
\figurenum{2} \epsscale{0.6}
\plotfiddle{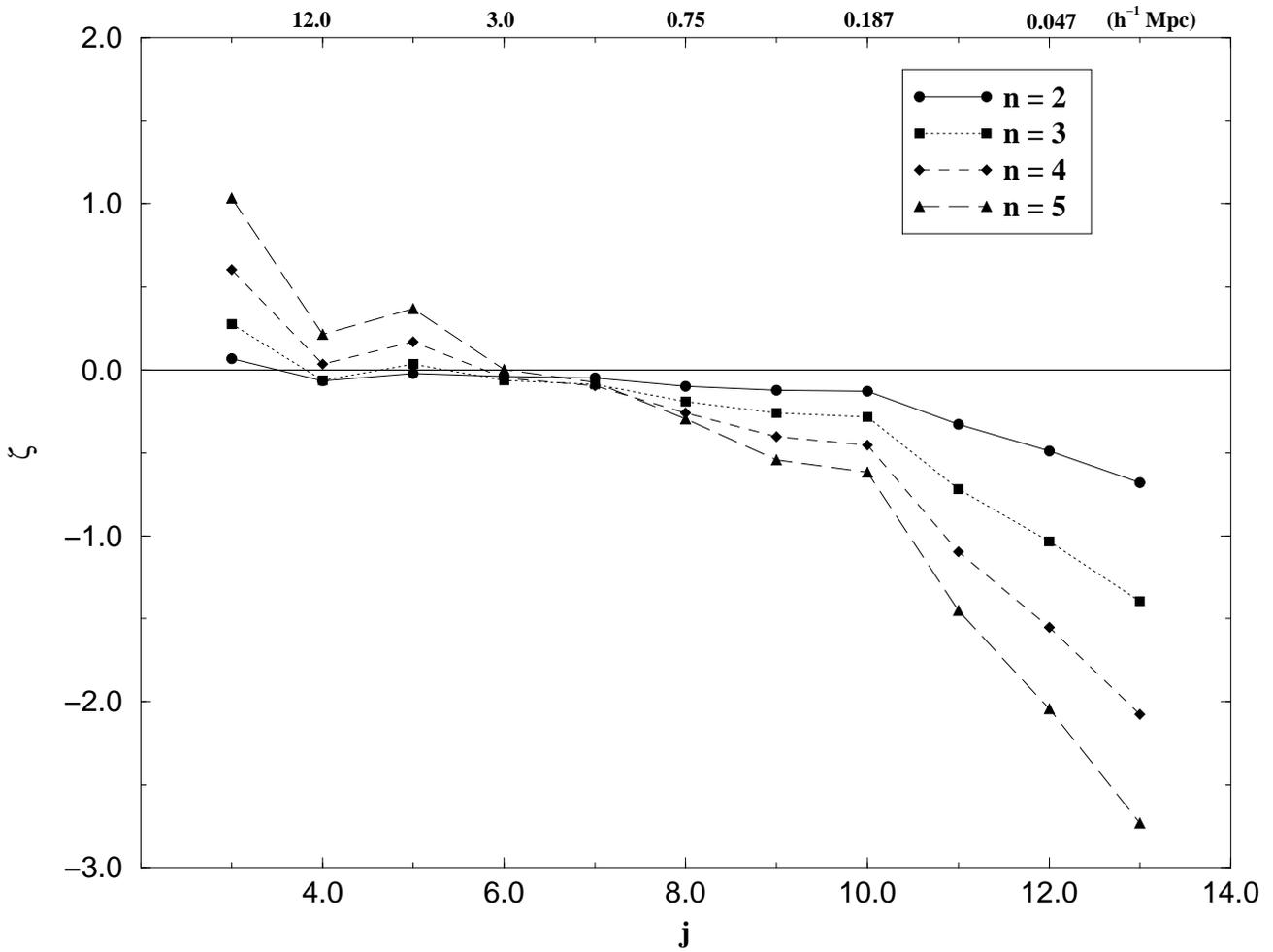}{20.cm}{270}{80}{80}{-310.}{440.}
\caption[]{The intermittent exponent $\zeta$ vs. $j$ of the
Ly$\alpha$ transmitted flux of QSO HS1700+64. The top scale is
the comoving scale, i.e., $189.84/2^j$ h$^{-1}$ Mpc.}
\end{figure}

\begin{figure}
\figurenum{3} \epsscale{0.6}
\plotfiddle{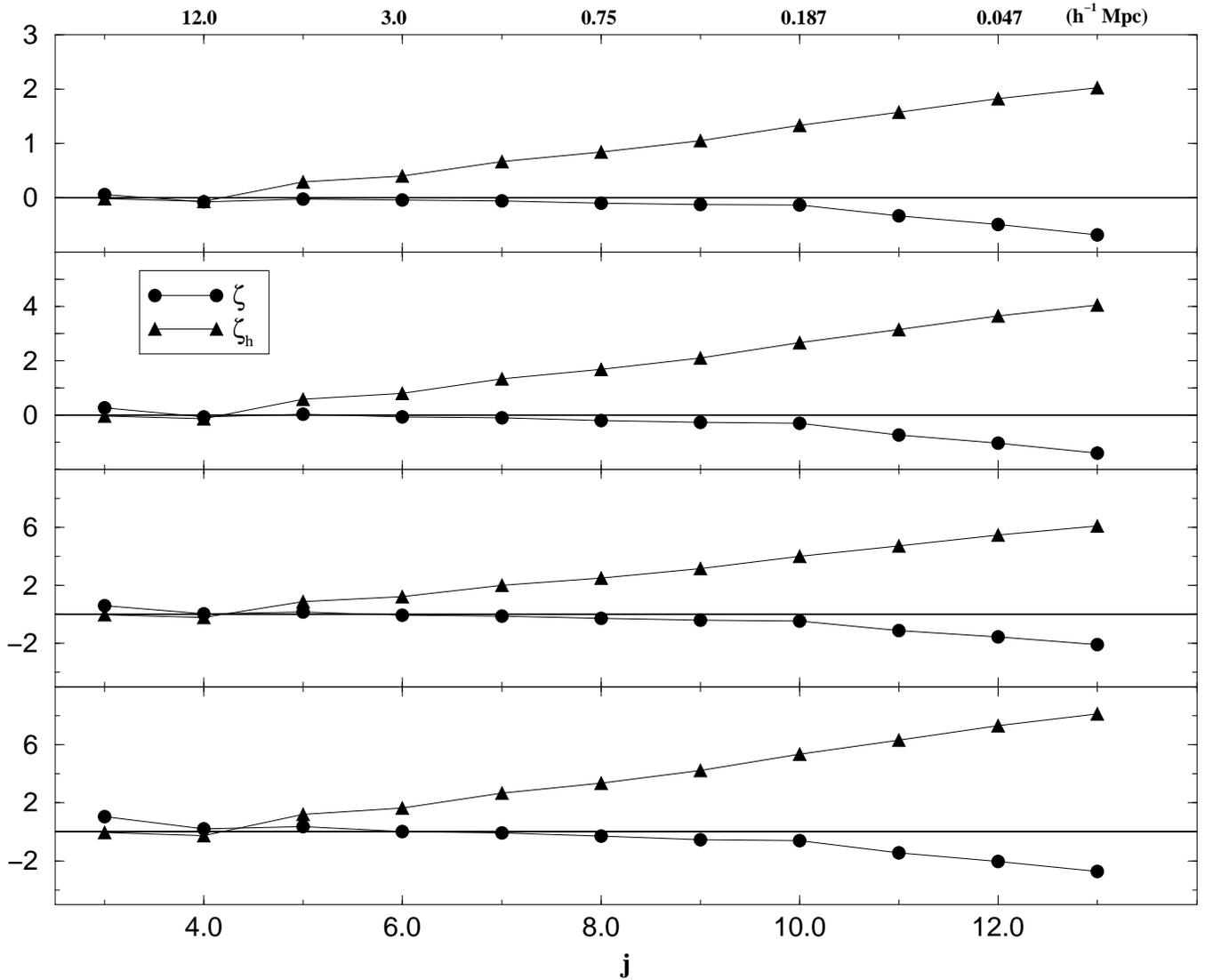}{20.cm}{270}{80}{80}{-310.}{440.}
\caption[]{The intermittent exponent $(\zeta_h)$ calculated by
the linked-pair approximation, $\zeta$ is the same as Fig. 2. The
top scale is the comoving scale, i.e., $189.84/2^j$ h$^{-1}$ Mpc.}
\end{figure}

\begin{figure}
\figurenum{4} \epsscale{0.6}
\plotfiddle{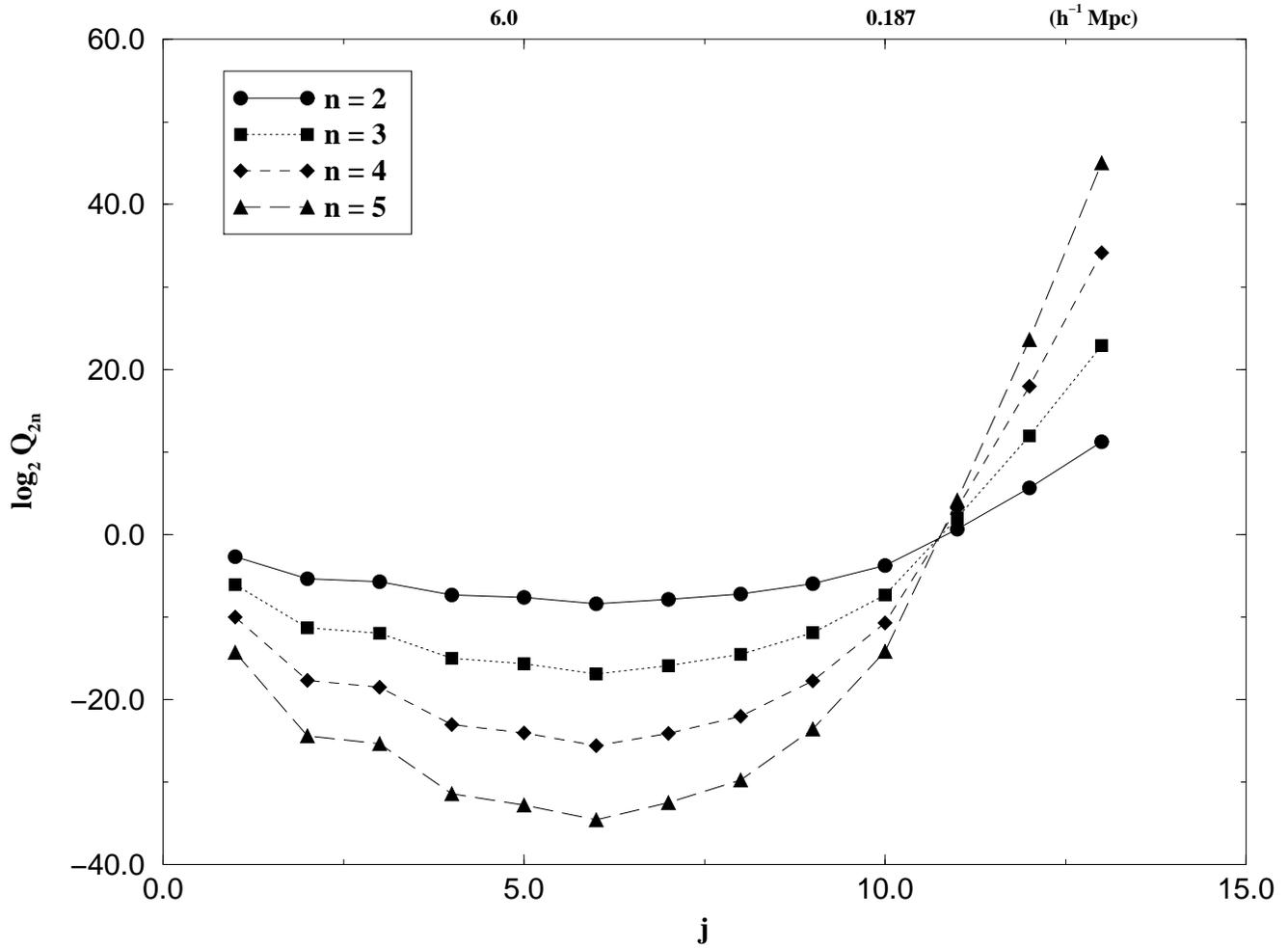}{20.cm}{270}{80}{80}{-310.}{440.}
\caption[]{The scale-dependence of $Q_{2n}$. The top scale is the
comoving scale, i.e., $189.84/2^j$ h$^{-1}$ Mpc.}
\end{figure}

\begin{figure}
\figurenum{5} \epsscale{0.6}
\plotfiddle{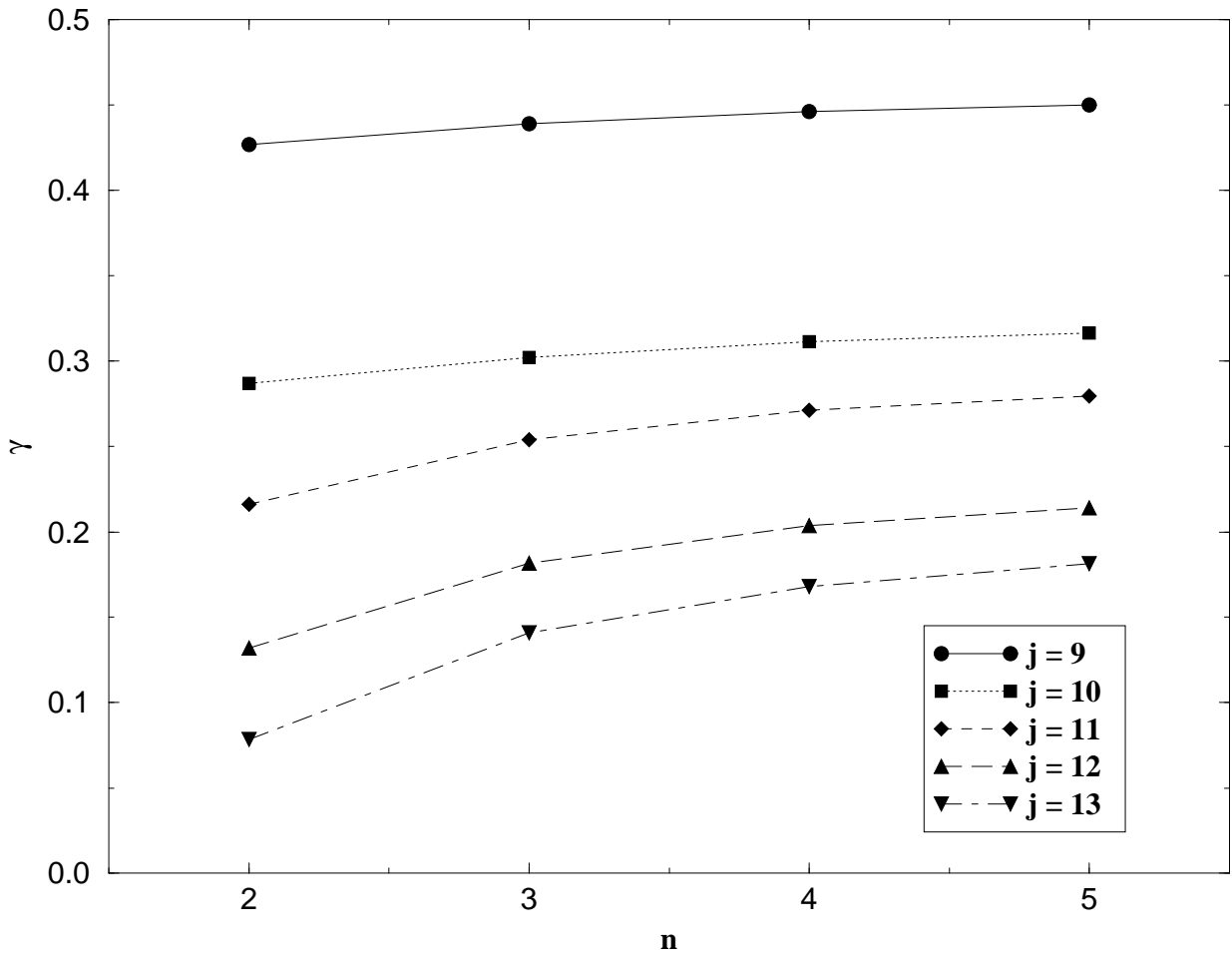}{20.cm}{270}{80}{80}{-310.}{440.}
\caption[]{The singular exponent $\gamma$ given by the asymptotic
behavior of eq.(45).}
\end{figure}

\end{document}